\documentclass[proceedings]{JHEP3}
\usepackage{amssymb}
\usepackage{eurosym}
\usepackage{amsfonts}
\usepackage{amsmath,cite}
\usepackage{epsfig}

\newbox\mybox

\newcommand\fverb{\setbox\mybox=\hbox\bgroup\verb}
\newcommand\fverbdo{\egroup\medskip\noindent\fbox{\unhbox\mybox}\ }
\newcommand\fverbit{\egroup\item[\fbox{\unhbox\mybox}]}
\conference{Exactly solvable quantum systems from point transformations}
\abstract{We demonstrate that complex point transformations can be used to construct non-Hermitian first integrals,
time-dependent Dyson maps and metric operators for non-Hermitian quantum systems. Initially we identify a point transformation as a map from an exactly solvable
time-independent system to an explicitly time-dependent non-Hermitian Hamiltonian system. Subsequently we employ the point transformation to construct the 
non-Hermitian time-dependent invariant for the latter system. Exploiting the fact that this invariant is pseudo-Hermitian, we
construct a corresponding Dyson map as the adjoint action from a non-Hermitian to a Hermitian invariant, thus obtaining solutions to the time-dependent Dyson 
and time-dependent quasi-Hermiticity equation together with solutions to the corresponding time-dependent Schr\"odinger equation.}

\title{Exactly solvable time-dependent non-Hermitian quantum systems from
point transformations}
\author{Andreas Fring and Rebecca Tenney \\
Department of Mathematics, City University London,\\
Northampton Square, London EC1V 0HB, UK\\
E-mail: a.fring@city.ac.uk, rebecca.tenney@city.ac.uk}

\input{tcilatex}
\begin{document}

\section{Introduction}

Point transformations are time-dependent canonical transformations used in
classical mechanics for a long time \cite{steeb1993invertible}. In this
context they are designed to extend standard transformations of the
configuration coordinates to the entire phase space of a system. In the
1950s they were utilized for the first time by DeWitt in quantum mechanics 
\cite{dewitt1952point,wolf1973point} in trying to settle the ambiguity
problem of operator ordering. This problem always emerges in the
quantization process of a theory when one seeks the quantum analogues for
classical expressions involving at least two factors whose mutual Poisson
bracket does not vanish.

In addition to solving fundamental conceptual issues in quantum mechanics,
point transformations have also been used to map simple exactly solvable
models to more complicated systems, including their solutions \cite%
{aldaya2011}, thus obtaining nontrivial information about the latter.
Exploiting the fact that point transformations preserve conserved quantities 
\cite{steeb1993invertible}, Zelaya and Rosas-Ortiz \cite{zelaya2020quantum}
recently showed that they may be employed to compute time-dependent
invariants or first integrals for Hermitian Hamiltonian systems. Here we
demonstrate that when complexifying these transformations they may also be
used to construct time-dependent invariants for non-Hermitian systems. The
explicit knowledge of these conserved quantities then allows to aid the
construction of time-dependent Dyson maps, and therefore metric operators,
by finding a similarity transformation. Proceeding in this manner one has
simplified the original problem of defining meaningful inner products as one
has circumvented solving the more complicated time-dependent Dyson equation
or time-dependent quasi-Hermiticity equation. Technically one has therefore
reduced the problem to finding the adjoint action that maps a non-Hermitian
invariant to a Hermitian one, in analogy to the familiar problem for
Hamiltonians with the difference that the map may become explicitly
time-dependent.

Our manuscript is organized as follows: In section 2 we outline the general
scheme in form of a four-step method leading to solutions of the
time-dependent Schr\"{o}dinger equation for an explicitly time-dependent
non-Hermitian Hamiltonian including a metric operator that ensures unitary
time-evolution between the obtained states. In section 3 we carry out the
first step in our procedure and set up a point transformation for various
Hermitian and non-Hermitian reference Hamiltonians leaving a number of
functions free that will be fixed in the next step when specifying a
concrete non-Hermitian target Hamiltonian. In section 4 we take this target
Hamiltonian to be the time-dependent Swanson model. In the next steps we
construct an invariant for this model and subsequently a Dyson map and
metric operator. In section 5 we carry out the same steps for another
non-Hermitian target Hamiltonian, a time-dependent harmonic oscillator with
complex linear term. Our conclusions are stated in section 6.

\section{Invariants and Dyson maps from point transformations}

Our starting problem is having to make sense of a non-Hermitian explicitly
time-dependent Hamiltonian $H(x,t)\neq H^{\dagger }(x,t)$ satisfying the
time-dependent Schr\"{o}dinger equation (TDSE)%
\begin{equation}
H(x,t)\phi (x,t)=i\hbar \partial _{t}\phi (x,t).  \label{TDSE1}
\end{equation}%
Unlike as for Hermitian Hamiltonians we do not only have to solve equation (%
\ref{TDSE1}) for the wavefunction $\phi (x,t)$, but we also have to find a
suitable time-dependent metric operator $\rho (t)$ for these solutions to
become physically meaningful in a well-defined inner product $\left\langle
\cdot \right. \left\vert \cdot \right\rangle _{\rho (t)}:=$ $\left\langle
\cdot \right. \left\vert \rho (t)\cdot \right\rangle $ \cite%
{CA,time1,time6,BilaAd,time7,fringmoussa,AndTom1,maamache2017pseudo,most2018en,BeckyAnd2}%
, similarly to the time-independent scenario \cite{BB,Alirev,PTbook}. In
principle one has to solve for this purpose the time-dependent Dyson
equation or the time-dependent quasi-Hermiticity equation 
\begin{equation}
h(t)=\eta (t)H(t)\eta ^{-1}(t)+i\hbar \partial _{t}\eta (t)\eta (t)^{-1},~~\
H^{\dagger }(t)=\rho (t)H(t)\rho ^{-1}(t)+i\hbar \partial _{t}\rho (t)\rho
^{-1}(t),  \label{TDDE}
\end{equation}%
respectively, for $\eta (t)$ or $\rho (t)=\eta ^{\dagger }(t)\eta (t)$. Here 
$h(t)=h^{\dagger }(t)$ is the time-dependent Hermitian counterpart to $H(t)$%
. While in many cases this is achievable, the construction of $\eta (t)$ and 
$\rho (t)$ is technically involved as demonstrated in \cite%
{BilaAd,time7,fringmoussa,fringmoussa2,AndTom1,maamache2017pseudo,most2018en,BeckyAnd2}%
.

The main purpose of this paper is to present an alternative approach to
finding $\rho (t)$ and $\eta (t)$ by exploiting \emph{point transformations}
and first integrals. As a starting point one assumes that there exists an
exactly solvable time-independent reference Hamiltonian $H_{0}(\chi )$
satisfying the one-dimensional TDSE 
\begin{equation}
H_{0}(\chi )\psi (\chi ,\tau )=i\hbar \partial _{\tau }\psi (\chi ,\tau ),
\label{TDSE2}
\end{equation}%
with $\chi $ denoting the coordinate and $\tau $ the time in this system.
One may then relate (\ref{TDSE2}) to the first TDSE (\ref{TDSE1}) by means
of a complex point transformation 
\begin{equation}
\Gamma :H_{0}\text{-TDSE}\rightarrow H\text{-TDSE, \ \ \ \ \ \ }[\chi ,\tau
,\psi (\chi ,\tau )]\mapsto \left[ x,t,\phi (x,t)\right] .  \label{gamma}
\end{equation}%
Here $\psi $ and $\phi $ are understood to be implicit functions of $(\chi $,%
$\tau )$ and $(x$,$t)$, respectively, defined by the equations (\ref{TDSE1})
and (\ref{TDSE2}). The variables $\chi $, $\tau $, $\psi $ are treated in
general as functions of the independent variables $x$, $t$, $\phi $ as%
\begin{equation}
\chi =P(x,t,\phi ),\text{\qquad }\tau =Q(x,t,\phi ),\qquad \psi =R(x,t,\phi
).  \label{genpoint}
\end{equation}%
In practice one may relax some of the $(x,t,\phi )$-dependences of the
functions $P$, $Q$, $R$ or is even forced to do so for concrete systems. We
refer here to $H_{0}$ and $H$, as \emph{reference} and \emph{target
Hamiltonians} respectively, not to be confused with their corresponding
Hermitian counterparts $h_{0}$ and $h$ in case they are non-Hermitian.

Having identified the point transformation $\Gamma $ on the level of the
TDSEs one may subsequently apply it exclusively to the time-independent
Hamiltonian $H_{0}(\chi )$ as 
\begin{equation}
\Gamma :H_{0}(\chi )\rightarrow I_{H}(x,t).  \label{GammaH0}
\end{equation}%
Since real point transformations preserve conserved quantities \cite%
{steeb1993invertible}, and $I_{H}(x,t)$ acquired a time-dependence via the
point transformation $\Gamma $, it is suggestive to assume that also complex
point transformations have this property and that $I_{H}(x,t)$ is actually
the time-dependent conserved \emph{Lewis-Riesenfeld invariant} \cite{Lewis69}
for the non-Hermitian time-dependent Hamiltonian $H(x,t)$ in (\ref{TDSE1})
satisfying 
\begin{equation}
i\hbar \frac{dI_{H}}{dt}=i\hbar \partial _{t}I_{H}+[I_{H},H]=0.  \label{Inv}
\end{equation}%
Since $H$ is non-Hermitian, also its first integral, the invariant $I_{H}$,
must be non-Hermitian, which is evident from (\ref{Inv}).

\noindent As argued successfully in \cite%
{khantoul2017invariant,maamache2017pseudo,AndTom4,cen2019time} one can map
this non-Hermitian invariant $I_{H}$ to a Hermitian invariant $I_{h}$ by
means of a time-dependent similarity transformation $\eta (t)$ as 
\begin{equation}
\eta (t)I_{H}(t)\eta ^{-1}(t)=I_{h}(t).  \label{sim}
\end{equation}%
Remarkably the map $\eta (t)$ is indeed the Dyson map solving the first
equation in (\ref{TDDE}) and the Hermitian operator $I_{h}$ is the
Lewis-Riesenfeld invariant for the Hermitian time-dependent Hamiltonian $%
h(t) $, identified in (\ref{TDDE}), satisfying 
\begin{equation}
i\hbar \frac{dI_{h}}{dt}=i\hbar \partial _{t}I_{h}+[I_{h},h]=0.
\end{equation}%
The metric operator is the simply obtained as $\rho (t)=\eta ^{\dagger
}(t)\eta (t)$.

In summary, we have proposed a four step method that leads not only to the
solutions of the TDSE (\ref{TDSE1}), but also to explicit expressions for
the Dyson maps and metric operators. The first step consists of selecting a
suitable time-independent reference Hamiltonian $H_{0}(\chi )$, Hermitian or
non-Hermitian, and point transform it's corresponding TDSE (\ref{TDSE2}). In
the second step we fix the free parameters by matching the transformed TDSE
with a TDSE (\ref{TDSE1}) for a non-Hermitian target Hamiltonian $H(t)$,
hence identifying the point transformation $\Gamma $ by means of (\ref{gamma}%
). In the third step we obtain the invariant $I_{H}(t)$ by acting with $%
\Gamma $ on the time-independent reference Hamiltonian $H_{0}(\chi )$ and in
the fourth step we construct the Dyson map $\eta $ as a similarity
transformation by means of (\ref{sim}). In case the TDSE for $H_{0}(\chi )$
is solvable we obtain by construction also the solutions to the original
TDSE for $H(x,t)$. Here our main focus is, however, on the construction of $%
I_{H}$, $I_{h}$, $\eta $ and $\rho $. Let us now demonstrate how this four
step strategy is carried out for some concrete time-dependent non-Hermitian
Hamiltonian.

\section{Point transforming the reference Hamiltonian}

One of the simplest choices for an exactly solvable reference Hamiltonian $%
H_{0}(\chi )$ is the time-independent Hermitian harmonic oscillator%
\begin{equation}
H_{0}(\chi )=\frac{P^{2}}{2m}+\frac{1}{2}m\omega ^{2}\chi
^{2},~~~~~~~m,\omega \in \mathbb{R}\text{.}  \label{HO}
\end{equation}%
First we identify the point transformation of $H_{0}(\chi )$ in general
terms. Expressing the momentum operator $P$ in the position representation $%
P=-i\hbar \partial _{\chi }$, we act with the point transformation $\Gamma $
on the TDSE (\ref{TDSE2}). Simplifying the general functional dependence as
stated in (\ref{genpoint}) to 
\begin{equation}
\chi =\chi (x,t),\qquad \tau =\tau (t),\qquad \psi =A(x,t)\phi (x,t),
\label{assum}
\end{equation}%
we convert all terms in the TDSE from the $(\chi ,\tau ,\psi )$ to the $%
(x,t,\phi )$-variables, hence obtaining the point transformed differential
equation%
\begin{equation}
i\hbar \phi _{t}+\frac{\hbar ^{2}}{2m}\frac{\tau _{t}}{\chi _{x}^{2}}\phi
_{xx}+B_{0}(x,t)\phi _{x}-V_{0}(x,t)\phi =0,  \label{trans}
\end{equation}%
with%
\begin{eqnarray}
B_{0}(x,t) &=&-i\hbar \frac{\chi _{t}}{\chi _{x}}+\frac{\hbar ^{2}}{2m}\frac{%
\tau _{t}}{\chi _{x}^{2}}\left( 2\frac{A_{x}}{A}-\frac{\chi _{xx}}{\chi _{x}}%
\right) ,  \label{B0} \\
V_{0}(x,t) &=&\frac{1}{2}m\tau _{t}\chi ^{2}\omega ^{2}-i\hbar \left( \frac{%
A_{t}}{A}-\frac{A_{x}\chi _{t}}{A\chi _{x}}\right) -\frac{\hbar ^{2}}{2m}%
\frac{\tau _{t}}{\chi _{x}^{2}}\left( \frac{A_{xx}}{A}-\frac{A_{x}\chi _{xx}%
}{A\chi _{x}}\right) .  \label{V0}
\end{eqnarray}%
This form of equation (\ref{trans}) agrees with the previously derived
expression in \cite{zelaya2020quantum}, where also more details of the
computation can be found. However, we allow for a major difference by
admitting the potential $V_{0}$ of the target Hamiltonian to be complex. The
first two assumptions in (\ref{assum}) on the functional dependence when
compared to the most general dependence $\chi (x,t,\phi )$, $\tau (x,t,\phi
) $ are made for convenience to simplify the calculation. The last
factorization property of $\psi $ in (\ref{assum}) is already made in
anticipation on the structure of the target differential equation. Since the
TDSE is a linear equation in the fields it does not contain a $\phi _{x}^{2}$
term so that $\psi _{\phi \phi }=0$. Hence the linear dependence of $\psi $
on $\phi $.

The reference Hamiltonian is a choice and in order to allow for comparison
we shall explore here some further simple options 
\begin{align}
H_{0}^{(1)}(\chi )& =\frac{P^{2}}{2m}  \label{h1} \\
H_{0}^{(2)}(\chi )& =H_{0}(\chi )+a\chi ,\qquad \ a\in \mathbb{R}, \\
H_{0}^{(3)}(\chi )& =H_{0}(\chi )+ib\chi ,\qquad b\in \mathbb{R},  \label{h3}
\\
H_{0}^{(4)}(\chi )& =H_{0}(\chi )+a\{\chi ,P\}.  \label{h4}
\end{align}%
We note that the reference Hamiltonian does not have to be Hermitian. Then
the general form of the point transformed differential equation (\ref{trans}%
) associated with each these reference Hamiltonians remains the same, yet
the explicit forms of $B_{0}(x,t)$ (\ref{B0}) and $V_{0}(x,t)$ (\ref{V0})
differ. For the choices (\ref{h1})-(\ref{h4}) we obtain 
\begin{align}
B_{1}(x,t)& =B_{0}(x,t),\quad & V_{1}(x,t)& =V_{0}(x,t)-\frac{1}{2}m\omega
^{2}\chi ^{2}\tau _{t},  \label{b1} \\
B_{2}(x,t)& =B_{0}(x,t),\quad & V_{2}(x,t)& =V_{0}(x,t)+a\chi \tau _{t}, \\
B_{3}(x,t)& =B_{0}(x,t),\quad & V_{3}(x,t)& =V_{0}(x,t)+ib\chi \tau _{t},
\label{b3} \\
B_{4}(x,t)& =B_{0}(x,t)+\frac{2ia\hbar \chi \tau _{t}}{\chi _{x}},\quad & 
V_{4}(x,t)& =V_{0}(x,t)-\frac{2ia\chi \hbar A_{x}\tau _{t}}{A\chi _{x}}%
-ia\hbar \tau _{t}.  \label{b4}
\end{align}%
In order to proceed to the second step in the procedure we need to select a
specific target Hamiltonian.

\section{The time-dependent Swanson model as target Hamiltonian}

As a concrete example for a target Hamiltonian we consider here a prototype
non-Hermitian Hamiltonian system, the time-dependent version of the Swanson
Hamiltonian \cite{Swanson}. In terms of bosonic creation $a$ and
annihilation operators $a^{\dagger }$, its time-dependent version is usually
written in the form%
\begin{equation}
\tilde{H}_{S}(t)=\omega (t)\left( a^{\dagger }a+1/2\right) +\tilde{\alpha}%
(t)a^{2}+\tilde{\beta}(t)\left( a^{\dagger }\right) ^{2},  \label{Swan}
\end{equation}%
which is clearly non-Hermitian when $\tilde{\alpha}\neq \tilde{\beta}^{\ast }
$. Dyson maps for the time-independent and time-dependent version were found
in \cite{MGH,bagarello2017pseudo} and \cite{fringmoussa2}, respectively. In
order to apply the point transformations it is more convenient to convert
the Hamiltonian into coordinate and momentum variables $x$, $p$, which is
easily achieved. Using the standard representations $a=(x+ip)/2$ and $%
a^{\dagger }=(x-ip)/2$ we obtain 
\begin{equation}
\tilde{H}_{S}(t)=\frac{1}{2}\left[ \omega (t)-\tilde{\alpha}(t)-\tilde{\beta}%
(t)\right] p^{2}+\frac{1}{2}\left[ \omega (t)+\tilde{\alpha}(t)+\tilde{\beta}%
(t)\right] x^{2}+\frac{i}{2}\left[ \tilde{\alpha}(t)-\tilde{\beta}(t)\right]
\{x,p\}+\frac{\omega (t)}{2}.
\end{equation}%
Expressing the time-dependent function $\tilde{\alpha}(t)$, $\tilde{\beta}(t)
$, $\omega (t)$ in terms of new time-dependent functions $\alpha (t)$, $%
\Omega (t)$ and $M(t)$ as 
\begin{equation}
\tilde{\alpha}=\frac{M\Omega ^{2}}{4}-\frac{1}{4M}+\alpha \text{,~~~}\tilde{%
\beta}=\frac{M\Omega ^{2}}{4}-\frac{1}{4M}-\alpha \text{,~~~}\omega =\frac{%
M\Omega ^{2}}{2}+\frac{1}{2M},  \label{abo}
\end{equation}%
the Hamiltonian is converted into the simpler form 
\begin{equation}
H_{S}(x,t):=\tilde{H}_{S}(t)-\frac{\omega (t)}{2}=\frac{p^{2}}{2M(t)}+\frac{%
M(t)}{2}\Omega ^{2}(t)x^{2}+i\alpha (t)\{x,p\},~~M,\Omega \in \mathbb{R}%
\text{, }\alpha \in \mathbb{C}\text{,}
\end{equation}%
which is evidently still non-Hermitian for $\alpha \neq 0$. The Swanson
Hamiltonian is $\mathcal{PT}$-symmetric for $\mathcal{PT}$: $x\rightarrow -x$%
, $p\rightarrow p$, $i\rightarrow -i$ and all time-dependent coefficient
functions transforming as $\mathcal{PT}$: $M,\Omega ,\alpha \rightarrow
M,\Omega ,\alpha $. Since $\alpha =\alpha _{r}+i\alpha _{i}$ is complex,
this requires requires $\mathcal{PT}$: $\alpha _{r}\rightarrow \alpha _{r}$, 
$\alpha _{i}\rightarrow -\alpha _{i}$. We notice here that the option $%
\alpha \in \mathbb{C}$, rather than $\alpha \in \mathbb{R}$, does not exist
in the time-independent case when one wishes to maintain the $\mathcal{PT}$%
-symmetry of the Hamiltonian.

We will explore here two versions of this target Hamiltonian. In one we keep
the mass time-independent by setting the time-dependent coefficient in the
kinetic energy term to a constant, $M(t)\rightarrow m$, and in the other
option we take the mass term to be generically time-dependent \cite%
{pedrosa1997exact,zelaya2020point}. Let us now identify the point
transformation $\Gamma $ according to (\ref{gamma}) for the specified pairs
of Hamiltonians.

\subsection{Point transformations $\Gamma _{i}^{S}$ from $H_{0}^{i}(\protect%
\chi )$ to $H_{S}(x,t)$}

\subsubsection{Point transformation $\Gamma _{0}^{S}:H_{0}(\protect\chi )$ $%
\rightarrow $ $H_{S}(x,t)$, time-independent mass \label{point1}}

Having specified the target Hamiltonian as $H_{S}(x,t)$, with $m$
time-independent, we express the corresponding TDSE (\ref{TDSE2}) in the
position representation as%
\begin{equation}
i\hbar \phi _{t}+\frac{\hbar ^{2}}{2m}\phi _{xx}-2\hbar \alpha (t)x\phi
_{x}-\hbar \alpha (t)\phi -\frac{1}{2}m\Omega (t)x^{2}\phi =0.  \label{mm}
\end{equation}%
Taking $H_{0}(\chi )$ as reference Hamiltonian, the direct comparison with
equation (\ref{trans}) leads to the three constraints%
\begin{equation}
\frac{\tau _{t}}{\chi _{x}^{2}}=1,~~~B_{0}(x,t)=-2\hbar \alpha
(t)x,~~~V_{0}(x,t)=\frac{1}{2}m\Omega (t)x^{2}+\hbar \alpha (t).
\label{const}
\end{equation}%
Apart from being a complex equation, the first constraint in (\ref{const})
is the same as the one found in \cite{zelaya2020quantum}, where it was
solved by%
\begin{equation}
\tau (t)=\int^{t}\frac{ds}{\sigma ^{2}(s)},\qquad \text{and\qquad }\chi
(x,t)=\frac{x+\gamma (t)}{\sigma (t)},
\end{equation}%
but now with $\gamma (t)$ and $\sigma (t)$ potentially being complex
functions. Using these expressions in the second constraint in (\ref{const})
yields the equation%
\begin{equation}
i\frac{\hbar }{m}\frac{A_{x}}{A}+\gamma _{t}+2i\alpha x-(x+\gamma )\frac{%
\sigma _{t}}{\sigma }=0,
\end{equation}%
which may be solved by%
\begin{equation}
A(x,t)=\exp \left\{ \frac{im}{\hbar }\left[ \left( \gamma _{t}-\gamma \frac{%
\sigma _{t}}{\sigma }\right) tx+\left( it\alpha -\frac{\sigma _{t}}{2\sigma }%
\right) x^{2}+\delta (t)\right] \right\} ,
\end{equation}%
where $\delta (t)$ is a complex valued function corresponding to the
integration constant in the $x$ integration. Proceeding with these
expressions to the third constraint in (\ref{const}) yields 
\begin{eqnarray}
&&-i\hbar \frac{\sigma _{t}}{2\sigma }-\frac{m}{2}\left( 2\gamma \gamma _{t}%
\frac{\sigma _{t}}{\sigma }+\gamma _{t}^{2}+\gamma ^{2}\frac{\sigma _{t}^{2}%
}{\sigma ^{2}}-\frac{\omega ^{2}\gamma ^{2}}{\sigma ^{4}}-2m\delta
_{t}\right)  \label{con3} \\
&&~~~~+\frac{m\gamma }{\sigma }\left[ \sigma _{tt}-\frac{\gamma _{tt}}{%
\gamma }\sigma -\frac{\omega ^{2}}{\sigma ^{3}}\right] x+\frac{m}{2\sigma }%
\left[ \sigma _{tt}-\left( 2i\alpha _{t}-4\alpha ^{2}-\Omega \right) \sigma -%
\frac{\omega ^{2}}{\sigma ^{3}}\right] x^{2}=0.~~~~~~  \notag
\end{eqnarray}%
The $x$-independent term in (\ref{con3}) vanishes for%
\begin{equation}
\delta (t)=\frac{\gamma }{2\sigma }\left( \sigma \gamma _{t}-\gamma \sigma
_{t}\right) -\frac{i\hbar }{2m}\log \sigma .
\end{equation}%
Furthermore, we recognize that the square brackets of the coefficient
functions for the $x$ and $x^{2}$ dependent terms amount both to the
ubiquitous nonlinear Ermakov-Pinney equation \cite{Ermakov,Pinney} with the
constraint%
\begin{equation}
\kappa (t):=\frac{\gamma _{tt}}{\gamma }=2i\alpha _{t}-4\alpha ^{2}-\Omega .
\end{equation}%
The general solution to this version of the Ermakov-Pinney (EP) equation, as
given by the coefficient functions, can be constructed in terms of the two
fundamental solutions $u(t)$ and $v(t)$ to the equations $\ddot{u}$ $+\kappa
(t)u=0$, $\ddot{v}+\kappa (t)v=0$ as%
\begin{equation}
\sigma (t)=\left( Au^{2}+Bv^{2}+2Cuv\right) ^{1/2},
\end{equation}%
where the constants $A$, $B$, $C~$are constrained as $C^{2}=AB-\omega ^{2}/W$
with Wronskian $W=u\dot{v}-v\dot{u}$. Given that $\kappa (t)$ is now
complex, the time $\tau $ and the coordinate $\chi $ inevitably become
complex, unless we take $\alpha _{t}=0$. As we see from (\ref{abo}) the
latter option still keeps all the coefficients time-dependent although in a
somewhat more restricted form.

\subsubsection{Point transformation $\hat{\Gamma}_{0}^{S}$ $:H_{0}(\protect%
\chi )$ $\rightarrow $ $H_{S}(x,t)$, time-dependent mass \label{point2}}

Let us now switch on the time-dependence in the mass so that we have to
compare the transformed equation (\ref{trans}) with 
\begin{equation}
i\hbar \phi _{t}+\frac{\hbar ^{2}}{2M(t)}\phi _{xx}-2\hbar \alpha (t)x\phi
_{x}-\hbar \alpha (t)\phi -\frac{1}{2}M(t)\Omega ^{2}(t)x^{2}\phi =0,
\end{equation}%
instead of (\ref{mm}). The direct comparison then changes the three
constraints (\ref{const}) into 
\begin{equation}
\frac{\tau _{t}}{m\chi _{x}^{2}}=\frac{1}{M(t)},\quad B(x,t)=-2\hbar \alpha
(t)x,\quad V(x,t)=\frac{1}{2}M(t)\Omega ^{2}(t)x^{2}+\hbar \alpha (t).
\label{eq:constraints}
\end{equation}%
Thus, also the first constraint in (\ref{eq:constraints}) differs now from
the one found in \cite{zelaya2020quantum} as a result of the introduction of
an explicit time-dependent mass. As we show next, this change from a
time-independent to a time-dependent mass permits us to keep the time $\tau $
and the coordinate $\chi $ to be real for more generic time-dependent
coefficient functions. Taking a general form for the mass as 
\begin{equation}
M(t)=m\sigma (t)^{n},  \label{MM}
\end{equation}%
allows us to easily to distinguish between the time-independent and
time-dependent cases, with the former recovered for $n=0$. The first
constraint in (\ref{eq:constraints}) is now solved by 
\begin{equation}
\tau (t)=\int^{t}\sigma (y)^{r}dy\qquad \text{and}\qquad \chi (x,t)=\frac{%
x+\gamma (t)}{\sigma (t)^{s}},  \label{eq:tx}
\end{equation}%
where we identify $n=-r-2s$. Using these expressions in the second
constraint in (\ref{eq:constraints}) yields the equation 
\begin{equation}
\sigma ^{r+2s}\frac{h}{m}\frac{A_{x}}{A}-i\gamma _{t}+is(x+\gamma )\frac{%
\sigma _{t}}{\sigma }+2\alpha x=0,
\end{equation}%
which may be solved by 
\begin{equation}
A(x,t)=\exp \left\{ \frac{im\sigma ^{-1-r-2s}}{\hbar }\left[ \left( \sigma
\gamma _{t}-s\gamma \sigma _{t}\right) x+\left( i\alpha \sigma -\frac{1}{2}%
s\sigma _{t}\right) x^{2}+\delta (t)\right] \right\} ,
\end{equation}%
where $\delta (t)$ is a complex valued function corresponding to the
integration constant in the $x$ integration. Proceeding with these
expressions to the third constraint in (\ref{eq:constraints}) yields 
\begin{align}
& -i\hbar \frac{q\sigma ^{1+r+2s}\sigma _{t}}{2}+\frac{m}{2}\left[
2(1+r+2s)\delta \sigma _{t}-2\sigma \delta _{t}+\sigma ^{2}\gamma
_{t}^{2}-2s\gamma \sigma \gamma _{t}\sigma _{t}+\gamma ^{2}\left(
s^{2}\sigma _{t}^{2}-\omega ^{2}\sigma ^{2+2s}\right) \right]  \notag
\label{eq:eq1} \\
& -m\left\{ \gamma \omega ^{2}\sigma ^{2r+2}-\sigma \lbrack r+2s]\gamma
_{t}\sigma _{t}+\gamma s\left[ (r+s+1)\sigma _{t}^{2}-\sigma \sigma _{tt}%
\right] +\sigma ^{2}\gamma _{tt}\right\} x+  \notag \\
& \frac{1}{2}m\left\{ 2i\alpha \sigma \lbrack r+2s]\sigma _{t}-2i\sigma
^{2}\alpha _{t}-s[r+s+1]\sigma _{t}^{2}+\sigma \left[ \sigma \left( 4\alpha
^{2}-\omega ^{2}\sigma ^{2r}+\Omega ^{2}\right) +s\sigma _{tt}\right]
\right\} x^{2}=0.
\end{align}%
The $x$-independent term in (\ref{eq:eq1}) vanishes for 
\begin{equation}
\delta (t)=\frac{\gamma }{2}\left( \sigma \gamma _{t}-s\gamma \sigma
_{t}\right) +\sigma ^{1+r+2s}\left( c_{1}-\frac{is\hbar }{2m}\log \sigma
\right) ,
\end{equation}%
where $c_{1}$ is a constant. The term proportional to $x^{2}$ in (\ref%
{eq:eq1}) is a non-linear second order differential equation in $\sigma $.
To ensure that $\sigma $ is real, hence our space-time is real, we set the
imaginary term to be equal to zero 
\begin{equation}
\alpha _{r}\left[ (r+2s)\sigma _{t}-4\sigma \alpha _{i}\right] -\sigma
(\alpha _{r})_{t}=0.
\end{equation}%
This equation is satisfied for 
\begin{equation}
\alpha _{i}=\frac{1}{4}\partial _{t}\ln \left( \frac{\sigma ^{r+2s}}{\alpha
_{r}}\right) .  \label{eq:ar}
\end{equation}%
We notice from here that since $\alpha _{i}\propto \partial _{t}$ it does
indeed transform as $\alpha _{i}\rightarrow -\alpha _{i}$ under $\mathcal{PT}
$ as is required for $H_{S}$ to be $\mathcal{PT}$-symmetric. The terms
proportional to $x^{2}$ and $x$ vanish for 
\begin{equation}
\sigma _{tt}=\sigma \left[ \frac{2\alpha _{r}\left( 2\Omega ^{2}\alpha
_{r}+8\alpha _{r}^{3}+(\alpha _{r})_{tt}\right) -3\left( \alpha _{r}\right)
_{t}^{2}}{2r\alpha _{r}^{2}}\right] +\frac{\left( \frac{r}{2}+1\right)
\sigma _{t}^{2}}{\sigma }-\frac{2\omega ^{2}\sigma ^{2r+1}}{r},
\label{eq:sig}
\end{equation}%
and 
\begin{eqnarray}
\gamma _{tt} &=&\frac{\gamma }{2r}\left( \frac{s\left( 16\alpha
_{r}^{4}-3\left( \alpha _{r}\right) _{t}^{2}+2\alpha _{r}(\alpha
_{r})_{tt}\right) }{\alpha _{r}^{2}}+4s\Omega ^{2}-\frac{(r+2s)\left(
2\omega ^{2}\sigma ^{2r+2}+rs\sigma _{t}^{2}\right) }{\sigma ^{2}}\right)
\label{eq:gam} \\
&&+\frac{(r+2s)\gamma _{t}\sigma _{t}}{\sigma }  \notag
\end{eqnarray}%
respectively. These equations can be reduced to solvable ones for specific
choices of $r$, $s$, $\alpha _{i}$, $\alpha _{r}$ and $\gamma $. We discuss
now some special choices.

\subsubsection*{\protect\underline{$\mathbf{\protect\alpha _{i}=0}$}}

Setting now $\alpha _{i}=0$, we can solve directly for $\alpha _{r}$ in (\ref%
{eq:ar}), obtaining 
\begin{equation}
\alpha _{r}=c_{2}\sigma ^{r+2s}.
\end{equation}%
Taking the mass to be time-independent and hence $\alpha $ to be
time-independent by setting $r=-2s$ and $s=1$, equations (\ref{eq:sig}) and (%
\ref{eq:gam}) reduce to 
\begin{equation}
\sigma _{tt}=-4c^{2}\sigma +\frac{\omega ^{2}}{\sigma ^{3}}-\sigma \Omega
^{2}\qquad \text{and}\qquad \gamma _{tt}=-\gamma \left( 4c^{2}+\Omega
^{2}\right) ,
\end{equation}%
respectively. Both of these equations are solvable, with the first being the
nonlinear Ermakov-Pinney equation \cite{Ermakov,Pinney}. Another interesting
choice is to take $r=-s-1$ with $s=-1$, in doing so we end up with 
\begin{equation}
\sigma _{tt}=\frac{4c^{2}}{\sigma ^{3}}-\sigma \omega ^{2}+\sigma \Omega
^{2}\qquad \text{and}\qquad \gamma _{tt}=-\gamma \left( \frac{4c^{2}}{\sigma
^{4}}+\Omega ^{2}\right) -\frac{2\gamma _{t}\sigma _{t}}{\sigma }
\end{equation}%
where again the first equation is a version of the nonlinear EP equation.
However, now the Ermakov-Pinney equation is real without any restrictions on 
$\alpha (t)$, so that also the time $\tau $ and the coordinate $\chi $ are
real. The second equation is a damped harmonic oscillator equation, which we
may solve or simply take the integration constant $\gamma $ to be zero.

\subsubsection*{\protect\underline{$\mathbf{\protect\gamma =0}$}}

Instead, setting $\gamma =0$ and parametrizing 
\begin{equation}
\alpha _{r}=\sigma ^{-2-r},
\end{equation}%
reduces equation (\ref{eq:sig}) to 
\begin{equation}
\sigma _{tt}=\frac{-\omega ^{2}\sigma ^{2r+1}+4\sigma ^{-2r-3}+\sigma \Omega
^{2}}{r+1},  \label{eq:epn}
\end{equation}%
with $\alpha $ now being being genuinely complex 
\begin{equation}
\alpha =\alpha _{r}-i\frac{r+s+1}{2}\partial _{t}\ln \left( \sigma \right) .
\end{equation}%
Choosing $r=0$ or $r=-2$ results in equation (\ref{eq:epn}) being the
respective EP equations given by 
\begin{equation}
\sigma _{tt}=\frac{4}{\sigma ^{3}}+\sigma \left( \Omega ^{2}-\omega
^{2}\right) ,~~~\text{or~~~}\sigma _{tt}=\frac{\omega ^{2}}{\sigma ^{3}}%
-\sigma \left( \Omega ^{2}+4\right) .
\end{equation}%
As we have taken $\gamma =0$ we do not need to select a concrete value for $%
s $.

When setting $r=-2$ we do not need to choose a concrete form for $\alpha
_{r} $, as in this case equation (\ref{eq:sig}) reduces to the
Ermakov-Pinney equation 
\begin{equation}
\sigma _{tt}=\frac{\omega ^{2}}{\sigma ^{3}}-f(t)\sigma ,~~\text{with~~}%
f=4\alpha _{r}^{2}-\frac{3\left( \alpha _{r}\right) _{t}^{2}}{4\alpha
_{r}^{2}}+\frac{(\alpha _{r})_{tt}}{2\alpha _{i}}+\Omega ^{2}.
\end{equation}

\subsubsection*{\protect\underline{$\mathbf{\protect\gamma \neq 0}$}}

When $\gamma \neq 0$, we still have the same parametrization of $\alpha _{r}$
and choices for $r$ as in the previous section, but we now have to restrict $%
s$ so that equation (\ref{eq:sig}) becomes solvable. For instance, when $%
r=-2 $ and $s=1$, we obtain 
\begin{equation}
\gamma _{tt}=-\gamma (4+\Omega ^{2}),
\end{equation}%
which is solvable.

\subsubsection{Point transformations $\hat{\Gamma}%
_{1,2,4}^{S}:H_{0}^{(1,2,4)}(\protect\chi )$ $\rightarrow $ $H_{S}(x,t)$,
time-dependent mass \label{point3}}

Let us next explore the point transformations that result when changing the
reference Hamiltonian, but keeping the target Hamiltonian to be $H_{S}(x,t)$
with time-dependent mass. Considering now the second constraint in (\ref%
{const}) together with (\ref{b1})-(\ref{b4}) we can identify the fields $%
A_{i}(x,t)$ for the reference Hamiltonians (\ref{h1})-(\ref{h4}). Solving
the constraints we find 
\begin{equation}
A_{1}(x,t)=A_{2}(x,t)=A(x,t),~~~\ A_{4}(x,t)=A(x,t)\exp \left[ \frac{%
am\sigma ^{-2s}}{i\hbar }\left( 2\gamma x+x^{2}\right) \right] ,
\end{equation}%
such that the $A_{i}(x,t)$ are identical for the same $B_{i}(x,t)$. Solving
next the third constraint in (\ref{const}) for (\ref{h1})-(\ref{h4}) we
notice that we always require (\ref{eq:ar}) to hold in order to ensure that
space-time remains real. In contrast, the other time-dependent functional
coefficient $\delta $ and the constraining equations for $\sigma $ and $%
\gamma $ vary for each reference Hamiltonians. We obtain%
\begin{align*}
H_{0}^{(1)}& :\quad \delta _{0}^{(1)}=\delta ,\quad \sigma
_{tt}^{(1)}=\sigma _{tt}+\frac{2\omega ^{2}\sigma ^{1+2r}}{r},\quad \gamma
_{tt}^{(1)}=\gamma _{tt}+\frac{(r+2s)\omega ^{2}\gamma \sigma ^{2r}}{r}, \\
H_{0}^{(2)}& :\quad \delta _{0}^{(2)}=\delta -\sigma ^{1+r+2s}\frac{a}{2m}%
\int^{t}\gamma \sigma ^{r-s},\quad \sigma ^{(2)}=\sigma ,\quad \gamma
_{tt}^{(2)}=\gamma _{tt}-\frac{a\sigma ^{2r+s}}{m}, \\
H_{0}^{(4)}& :\quad \delta _{0}^{(4)}=\delta +2a\sigma
^{1+r+2s}\int^{t}\gamma \sigma ^{-1-2s}(s\gamma \sigma _{t}-\sigma \gamma
_{t}),\quad \sigma _{tt}^{(4)}=\sigma _{tt}+\frac{8a^{2}\sigma ^{1+2r}}{r}%
,\quad \\
& ~~~~~\gamma _{tt}^{(4)}=\gamma _{tt}+\frac{4a^{2}(r+2s)\gamma \sigma ^{2r}%
}{r}.
\end{align*}
Here we understand that $\sigma _{tt}$ and $\gamma _{tt}$ are to be replaced
by the right hand sides of equations (\ref{eq:sig}) and (\ref{eq:gam}),
respectively.

\subsection{Non-Hermitian invariants from $\Gamma _{i}^{S}$}

Having constructed the various point transformations $\Gamma _{i}^{j}$ that
relate the TDSEs (\ref{TDSE1}) and (\ref{TDSE2}) for $H^{j}(x,t)$ and $%
H_{0}^{i}(\chi )$, respectively, we proceed to the third step in our scheme
and employ the point transformations now to act on $H_{0}^{i}(\chi )$
exclusively, as specified in (\ref{GammaH0}). In this way we obtain directly
the invariant $I_{H}$ for the non-Hermitian Hamiltonian $H$.

\subsubsection{Non-Hermitian invariant from $\Gamma _{0}^{S}$,
time-independent mass}

Acting with $\Gamma _{0}^{S}$, as constructed in section \ref{point1}, on $%
H_{0}(\chi )$ we obtain the invariant%
\begin{eqnarray}
I_{H}(x,t) &=&\frac{\sigma ^{2}}{2m}p^{2}+m\left( \frac{\gamma \omega ^{2}}{%
\sigma ^{2}}+2i\alpha (\sigma ^{2}\gamma _{t}-\gamma \sigma \sigma
_{t})-\sigma \sigma _{t}\gamma _{t}+\gamma \sigma _{t}^{2}\right) x+\sigma
\left( \sigma \gamma _{t}-\gamma \sigma _{t}\right) p  \notag \\
&&+\frac{1}{2}\sigma \left[ 2i\alpha \sigma -\sigma _{t}\right] \{x,p\}+%
\frac{m}{2}\left[ \left( \sigma _{t}-2i\alpha \sigma \right) {}^{2}+\frac{%
\omega ^{2}}{\sigma ^{2}}\right] ~x^{2}  \label{I2} \\
&&+\frac{m}{2}\left( \frac{\gamma ^{2}\omega ^{2}}{\sigma ^{2}}+\gamma
^{2}\sigma _{t}^{2}+\sigma ^{2}\gamma _{t}^{2}-2\gamma \gamma _{t}\sigma
\sigma _{t}\right) .  \notag
\end{eqnarray}%
We verified that the expression for $I_{H}$ in (\ref{I2}) does indeed
satisfy the Lewis-Riesenfeld equation (\ref{Inv}). Thus $I_{H}(x,t)$ is the
non-Hermitian invariant or first integral for the non-Hermitian Hamiltonian $%
H(x,t)$. We stress that the invariant has been obtained by a direct
calculation and did not involve any assumption or guess work on the general
form of the invariant, which one usually has to make when solving (\ref{Inv}%
) directly.

\subsubsection{Non-Hermitian invariant from $\hat{\Gamma}_{0}^{S}$,
time-dependent mass}

Similarly acting with $\hat{\Gamma}_{0}^{S}$, as constructed in section \ref%
{point2}, on $H_{0}(\chi )$ we obtain the invariant%
\begin{align}
\hat{I}_{H}(x,t)& =\frac{\sigma ^{2s}}{2m}p^{2}+\left( \sigma ^{-r}\gamma
_{t}-\gamma s\sigma ^{-r-1}\sigma _{t}\right) p+\frac{4i\sigma \alpha
_{r}^{2}+r\alpha _{r}\sigma _{t}-\sigma {\alpha _{r}}_{t}}{4\alpha
_{r}\sigma ^{r+1}}\{x,p\}  \notag \\
& +\frac{4m\omega ^{2}\alpha _{r}^{2}\sigma ^{2r+2}-m\left( 4\sigma \alpha
_{r}^{2}-ir\alpha _{r}\sigma _{t}+i\sigma {\alpha _{r}}_{t}\right) {}^{2}}{%
8\alpha _{r}^{2}\sigma ^{2(r+s+1)}}x^{2}  \notag \\
& +\frac{2\gamma m\omega ^{2}\alpha _{r}\sigma ^{2r+2}+m\left( \sigma \gamma
_{t}-\gamma s\sigma _{t}\right) \left( 4i\sigma \alpha _{r}^{2}+r\alpha
_{r}\sigma _{t}-\sigma {\alpha _{r}}_{t}\right) }{2\alpha _{r}\sigma
^{2(r+s+1)}}x  \notag \\
& +\frac{1}{2}m\sigma ^{-2(r+s+1)}\left[ \gamma ^{2}\omega ^{2}\sigma
^{2r+2}+\left( \sigma \gamma _{t}-\gamma s\sigma _{t}\right) {}^{2}\right]
\end{align}%
Once more we convince ourselves that $\hat{I}_{H}(x,t)$ does indeed satisfy (%
\ref{Inv}).

\subsubsection{Non-Hermitian invariant from $\hat{\Gamma}_{1,2,4}^{S}$,
time-dependent mass}

The action of $\hat{\Gamma}_{1,2,4}^{S}$ from section \ref{point3} on $%
H_{0}^{(1,2,4)}(\chi )$ yields the invariants%
\begin{align}
I_{H}^{(1)}(x,t)& =\frac{\sigma ^{2s}}{2m}p^{2}+\left( \sigma ^{-r}\gamma
_{t}-\gamma s\sigma ^{-r-1}\sigma _{t}\right) p+\frac{4i\sigma \alpha
_{r}^{2}+r\alpha _{r}\sigma _{t}-\sigma {\alpha _{r}}_{t}}{4\alpha
_{r}\sigma ^{r+1}}\{x,p\}  \notag \\
& -\frac{m\left( 4\sigma \alpha _{r}^{2}-ir\alpha _{r}\sigma _{t}+i\sigma {%
\alpha _{r}}_{t}\right) {}^{2}}{8\alpha _{r}^{2}\sigma ^{2(r+s+1)}}x^{2} 
\notag \\
& +\frac{m\left( \sigma \gamma _{t}-\gamma s\sigma _{t}\right) \left(
4i\sigma \alpha _{r}^{2}+r\alpha _{r}\sigma _{t}-\sigma {\alpha _{r}}%
_{t}\right) }{2\alpha _{r}\sigma ^{2(r+s+1)}}x  \notag \\
& +\frac{1}{2}m\sigma ^{-2(r+s+1)}\left( \sigma \gamma _{t}-\gamma s\sigma
_{t}\right) {}^{2},
\end{align}%
\begin{align}
I_{H}^{(2)}(x,t)& =\frac{\sigma ^{2s}}{2m}p^{2}+\left( \sigma ^{-r}\gamma
_{t}-\gamma s\sigma ^{-r-1}\sigma _{t}\right) p+\frac{\left( 4i\sigma \alpha
_{r}^{2}+r\alpha _{r}\sigma _{t}-\sigma {\alpha _{r}}_{t}\right) }{4\alpha
_{r}\sigma ^{r+1}}\{x,p\}  \notag \\
& +\frac{4m\omega ^{2}\alpha _{r}^{2}\sigma ^{2r+2}-m\left( 4\sigma \alpha
_{r}^{2}-ir\alpha _{r}\sigma _{t}+i\sigma {\alpha _{r}}_{t}\right) {}^{2}}{%
8\alpha _{r}^{2}\sigma ^{2(r+s+1)}}x^{2}  \notag \\
& +\frac{2a\alpha _{r}\sigma ^{2r+s+2}+2\gamma m\omega ^{2}\alpha _{r}\sigma
^{2r+2}+m\left( \sigma \gamma _{t}-\gamma s\sigma _{t}\right) \left(
4i\sigma \alpha _{r}^{2}+r\alpha _{r}\sigma _{t}-\sigma {\alpha _{r}}%
_{t}\right) }{2\alpha _{r}\sigma ^{2(r+s+1)}}x  \notag \\
& +\frac{1}{2\sigma ^{2(r+s+1)}}\gamma \sigma ^{2r+2}\left( 2a\sigma
^{s}+\gamma m\omega ^{2}\right) +m\left( \sigma \gamma _{t}-\gamma s\sigma
_{t}\right) {}^{2}
\end{align}%
and 
\begin{align}
I_{H}^{(4)}(x,t)& =\frac{\sigma ^{2s}}{2m}p^{2}+\left( \sigma ^{-r}\gamma
_{t}-\gamma s\sigma ^{-r-1}\sigma _{t}\right) p+\frac{4i\sigma \alpha
_{r}^{2}+r\alpha _{r}\sigma _{t}-\sigma {\alpha _{r}}_{t}}{4\alpha
_{r}\sigma ^{r+1}}\{x,p\}  \notag \\
& +\frac{-4m\left( 4a^{2}-\omega ^{2}\right) \alpha _{r}^{2}\sigma
^{2r+2}-m\left( 4\sigma \alpha _{r}^{2}-ir\alpha _{r}\sigma _{t}+i\sigma {%
\alpha _{r}}_{t}\right) {}^{2}}{8\alpha _{r}^{2}\sigma ^{2(r+s+1)}}x^{2} 
\notag \\
& +\frac{-2\gamma m\left( 4a^{2}-\omega ^{2}\right) \alpha _{r}\sigma
^{2r+2}+m\left( \sigma \gamma _{t}-\gamma s\sigma _{t}\right) \left(
4i\sigma \alpha _{r}^{2}+r\alpha _{r}\sigma _{t}-\sigma {\alpha _{r}}%
_{t}\right) }{2\alpha _{r}\sigma ^{2(r+s+1)}}x  \notag \\
& +\frac{1}{2\sigma ^{2(r+s+1)}}m\left[ \left( \sigma \gamma _{t}-\gamma
s\sigma _{t}\right) {}^{2}-\gamma ^{2}\left( 4a^{2}-\omega ^{2}\right)
\sigma ^{2r+2}\right] 
\end{align}%
Let us now compare the invariants obtained. First of all we notice that all
our invariants can be brought into the form 
\begin{equation}
I_{H}=a_{r}p^{2}+b_{r}p+\left( c_{r}+ic_{i}\right) \{x,p\}+\left(
d_{r}+id_{i}\right) x^{2}+\left( e_{r}+ie_{i}\right) x+f_{r},  \label{Ic}
\end{equation}%
where we abbreviated the complex time-dependent coefficient functions in $%
I_{H}$ and separate them into real and imaginary parts by denoting $%
x=x_{r}+ix_{i}$ with $x_{r},x_{i}\in \mathbb{R}$, $x\in \{a,b,c,d,e,f\}$.
When written in this form we notice a very peculiar property that for all of
our invariants the time-dependent coefficient functions are related to each
other as 
\begin{equation}
\frac{e_{i}}{2b_{r}}=\frac{d_{i}}{4c_{r}}=\frac{c_{i}}{2a_{r}}=\alpha
_{r}m\sigma ^{-r-2s}.  \label{sym}
\end{equation}%
As we will see in the next subsection this property is responsible for the
fact that all invariants lead to same Dyson map. Notice that when using the
conventions as in (\ref{Ic}) for the Hamiltonian $H_{S}(x,t)$ and employing
the same parameterization for $M(t)$ and $\alpha (t)$, the last relation
also holds for the coefficients in the Hamiltonian. We also note that if we
were to take $a\rightarrow ia$ in $H_{0}^{(4)}(\chi )$, with $%
c_{2}\rightarrow -c_{2}$, the associated invariant would still posses the
same properties as $a$ only appears squared in it. When comparing the
expressions for the invariants $I_{H}^{(i)}$ one needs to keep in mind that
the constraining equations also change with $i$.

\subsection{Dyson maps and metric operators}

We may now carry out the last step in our scheme and construct a Dyson map
by acting adjointly on the invariants $I_{H}$. We can verify that the Dyson
map constructed in \cite{fringmoussa2} does indeed map $I_{H}$ to a
Hermitian invariant. Alternatively, when utilizing the property (\ref{sym})
we also find a time-independent Dyson map 
\begin{equation}
\eta =\exp \left( -\alpha _{r}m\sigma ^{-r-2s}x^{2}\right) ,
\end{equation}%
with the associated time-dependent Hermitian invariant 
\begin{equation}
I_{h}=a_{r}p^{2}+b_{r}p+c_{r}\{x,p\}+\left( d_{r}+4m^{2}a_{r}\alpha
_{r}^{2}\sigma ^{-2r-4s}\right) x^{2}+e_{r}x+f_{r}.
\end{equation}%
The corresponding Hermitian Hamiltonian is computed to be 
\begin{equation}
h=\frac{\sigma ^{r+2s}}{2m}p^{2}+\left( 2m\alpha _{r}^{2}\sigma ^{-r-2s}+%
\frac{1}{2}m\sigma ^{-r-2s}\Omega ^{2}\right) x^{2}+\frac{1}{4}\partial
_{t}\ln \left( \frac{\sigma ^{r+2s}}{\alpha _{r}}\right) \{x,p\},
\end{equation}%
which is an extended version of the time-dependent harmonic oscillator with
time-dependent mass. For the special choice $\alpha _{r}=\sigma ^{r+2s}$ the
coefficient function $\alpha (t)$ becomes real, the Dyson map becomes
time-independent and $h$ reduces to the time-dependent harmonic oscillator.

\section{ The time-dependent harmonic oscillator with complex linear term as
target Hamiltonian}

To further illustrate the method and demonstrate the importance of the
choice of $H_{0}(\chi )$ we consider next the time-dependent harmonic
oscillator with a time-dependent complex linear term 
\begin{equation}
H_{CL}(x,t)=\frac{p^{2}}{2M(t)}+\frac{1}{2}M(t)\Omega ^{2}(t)x^{2}+i\beta
(t)x,\qquad M,\Omega ,\beta \in \mathbb{R},  \label{eq:newh}
\end{equation}%
which has been studied previously in \cite{timedep3,khantoul2017invariant}.
As a reference Hamiltonian we take now $H_{0}^{(3)}(\chi )$ as defined in (%
\ref{h3}). We have also considered $H_{0}(\chi )$ as a reference Hamiltonian
which leads to a point transformation that renders space-time to be complex.

\subsection{Point transformation $\Gamma _{3}^{CL}$ from $H_{0}^{(3)}(%
\protect\chi )$ to $H_{CL}(x,t)$}

We have already identified the equations for $B_{3}(x,t)$ and $V_{3}(x,t)$
for the reference Hamiltonian $H_{0}^{(3)}(\chi )$ in (\ref{b3}). Comparing
now with the time-dependent Schr\"{o}dinger equation for the target
Hamiltonian (\ref{TDSE2}) in the position representation we find the three
constraints 
\begin{equation}
\frac{\tau _{t}}{m\chi _{x}^{2}}=\frac{1}{M(t)},\qquad B(x,t)=0,\qquad
V(x,t)=\frac{1}{2}M(t)\Omega ^{2}(t)x^{2}+i\beta (t)x.  \label{eq:con2}
\end{equation}%
The first constraint in (\ref{eq:con2}) is solved in the same way as in
section \ref{point2}, i.e. by equations (\ref{eq:tx}), together with (\ref%
{MM}). Substituting these expressions into the second constraint in (\ref%
{eq:con2}) and then solving for the field $A(x,t)$ yields 
\begin{equation}
A(x,t)=\exp \left\{ \frac{im\sigma ^{-1-r-2s}}{\hbar }\left[ \left( \sigma
\gamma _{t}-s\gamma \sigma _{t}\right) x-\frac{1}{2}s\sigma _{t}x^{2}+\delta
(t)\right] \right\} ,
\end{equation}%
where $\delta (t)$ is a complex time-dependent function associated with the
integration carried out in $x$. Next we use all of our determined
expressions in the third constraint in (\ref{eq:con2}), obtaining 
\begin{align}
& 0=-m\left[ \omega ^{2}\sigma ^{2r+2}+s(r+s+1)\sigma _{t}^{2}-\sigma \left(
s\sigma _{tt}+\sigma \Omega ^{2}\right) \right] x^{2}+2i\sigma ^{r+2}\left(
\beta \sigma ^{2s}-b\sigma ^{r+s}\right) x  \notag  \label{eq:eq2} \\
& +2m\left[ \sigma (r+2s)\gamma _{t}\sigma _{t}+\gamma s\left( \sigma \sigma
_{tt}-(r+s+1)\sigma _{t}^{2}\right) -\sigma ^{2}\gamma _{tt}-\gamma \omega
^{2}\sigma ^{2r+2}\right] x-ihs\sigma _{t}\sigma ^{r+2s+1}  \notag \\
& +m\left\{ 2\sigma _{t}\left[ \delta (r+2s+1)-\gamma s\sigma \gamma _{t}%
\right] +\gamma ^{2}s^{2}\sigma _{t}^{2}+\sigma \left[ \sigma \gamma
_{t}^{2}-2\delta _{t}\right] \right\} -\gamma \sigma ^{2r+2}\left( \gamma
m\omega ^{2}+2ib\sigma ^{s}\right) .
\end{align}%
Firstly we notice that the $x$-dependent term in (\ref{eq:eq2}) contains an
imaginary term which would result in space-time becoming complex. However,
when setting 
\begin{equation}
\beta =b\sigma ^{r-s},
\end{equation}%
the imaginary term vanishes and space-time remains real. Secondly we find
that the $x$-independent terms in (\ref{eq:eq2}) vanishes for 
\begin{equation}
\delta (t)=\frac{\gamma }{2}\left( \sigma \gamma _{t}-s\gamma \sigma
_{t}\right) +\sigma ^{1+r+2s}\left( c_{1}-\frac{is\hbar }{2m}\log \sigma
-i\int^{t}\frac{b\gamma \sigma ^{r-s}}{m}\right) .
\end{equation}%
Finally, the remaining terms proportional to $x^{2}$ and $x$ result in the
two second order auxiliary differential equations 
\begin{equation}
\sigma _{tt}=\frac{\omega ^{2}\sigma ^{2r+2}-\sigma ^{2}\Omega ^{2}}{s\sigma 
}+\frac{(r+s+1)\sigma _{t}^{2}}{\sigma }\qquad \text{and}\qquad \gamma _{tt}=%
\frac{(r+2s)\gamma _{t}\sigma _{t}}{\sigma }-\gamma \Omega ^{2},  \label{sig}
\end{equation}%
respectively. As discussed in the previous section there are different
choices of $r$ and $s$ for which these equations reduce into versions with
known solutions. As before, we shall not select concrete values for $r$ and $%
s$ so we keep the derivation of the invariant and subsequent Dyson map as
general as possible.

\subsection{Non-Hermitian invariant from $\Gamma _{3}^{CL}$}

Acting with $\Gamma _{3}^{CL}$, as constructed in the previous section on $%
H_{0}^{(3)}(\chi )$ we obtain the invariant 
\begin{align}
I_{H}(x,t)& =\frac{\sigma ^{2s}}{2m}p^{2}+(\sigma ^{-r}\gamma _{t}-\gamma
s\sigma ^{-r-1}\sigma _{t})p-\frac{1}{2}s\sigma ^{-r-1}\sigma _{t}\{x,p\} 
\notag \\
& +\frac{1}{2}m\sigma ^{-2(r+s+1)}\left( \omega ^{2}\sigma
^{2r+2}+s^{2}\sigma _{t}^{2}\right) x^{2}  \notag \\
& \sigma ^{-2(r+s+1)}\left[ ms\sigma _{t}\left( \gamma s\sigma _{t}-\sigma
\gamma _{t}\right) +\sigma ^{2r+2}\left( \gamma m\omega ^{2}+ib\sigma
^{s}\right) \right] x  \notag \\
& +\frac{1}{2}\sigma ^{-2(r+s+1)}\left[ m\left( \sigma \gamma _{t}-\gamma
s\sigma _{t}\right) {}^{2}+\gamma \sigma ^{2r+2}\left( \gamma m\omega
^{2}+2ib\sigma ^{s}\right) \right] .  \label{IH}
\end{align}%
We have verified that this expression does indeed satisfy the
Lewis-Riesenfeld equation (\ref{Inv}).

\subsection{Time-dependent Dyson map and metric operator}

To determine the time-dependent Dyson map associated with the non-Hermitian
invariant (\ref{IH}) we use the following abbreviated version of the
invariant 
\begin{equation}
I_{H}=a_{r}p^{2}+b_{r}p+c_{r}\{x,p\}+d_{r}x^{2}+(e_{r}+ie_{i})x+f_{r}+if_{i},
\end{equation}%
using the same conventions as in (\ref{Ic}).

Making now the general Ansatz for the Dyson map 
\begin{equation}
\eta (t)=e^{\epsilon (t)p}e^{\lambda (t)x},\qquad \epsilon ,\lambda \in 
\mathbb{R},  \label{eq:dys}
\end{equation}%
we compute the adjoint action of the Dyson map on all the operators that
appear in the non-Hermitian invariant. We find that (\ref{eq:dys}) maps $%
I_{H}(x,t)$ indeed to a Hermitian counterpart when the following constraints
are satisfied 
\begin{equation}
\epsilon =\frac{a_{r}f_{i}}{a_{r}e_{r}-b_{r}c_{r}},\qquad \lambda =\frac{%
c_{r}\epsilon }{a_{r}},\qquad e_{i}=\frac{2(c_{r}^{2}-a_{r}d_{r})f_{i}}{%
b_{r}c_{r}-a_{r}e_{r}}.
\end{equation}%
The time-dependent functions from above do indeed satisfy these equations
and when using the explicit expressions for the coefficient functions from (%
\ref{IH}) the time-dependent Dyson map results to 
\begin{equation}
\eta (t)=\exp \left( \frac{b\sigma ^{s}}{m\omega ^{2}}p\right) \exp \left( -%
\frac{bs\sigma ^{-1-r-s}\sigma _{t}}{\omega ^{2}}x\right) ,
\end{equation}%
with $\sigma $ to be determined by the auxiliary equation (\ref{sig}). The
corresponding Hermitian invariant is computed to 
\begin{align}
I_{h}(x,t)& =\frac{\sigma ^{2s}}{2m}p^{2}+(\sigma ^{-r}\gamma _{t}-\gamma
s\sigma ^{-r-1}\sigma _{t})p-\frac{1}{2}s\sigma ^{-r-1}\sigma _{t}\{x,p\} 
\notag \\
& +\frac{1}{2}\frac{m}{\sigma ^{2(r+s+1)}}\left( \omega ^{2}\sigma
^{2r+2}+s^{2}\sigma _{t}^{2}\right) x^{2}+\frac{m}{\sigma ^{2(r+s+1)}}\left[
\gamma \omega ^{2}\sigma ^{2r+2}+s\sigma _{t}\left( \gamma s\sigma
_{t}-\sigma \gamma _{t}\right) \right] x  \notag \\
& +\frac{b^{2}+\gamma ^{2}m^{2}\omega ^{4}\sigma ^{-2s}}{2m\omega ^{2}}\frac{%
+m\left( \gamma ^{2}s^{2}\sigma _{t}^{2}+\sigma ^{2}\gamma _{t}^{2}\right) }{%
2\sigma ^{2(r+s+1)}}.
\end{align}%
Finally we use the Dyson map (\ref{eq:dys}) in the time-dependent Dyson
equation (\ref{TDDE}) to compute the corresponding Hermitian Hamiltonian as 
\begin{equation}
h(t)=\frac{\sigma ^{r+2s}}{2m}p^{2}+\frac{1}{2}m\sigma ^{-r-2s}\Omega
^{2}x^{2}+\frac{b^{2}\sigma ^{-r-2}\left( \sigma ^{2}\Omega ^{2}-s^{2}\sigma
_{t}^{2}\right) }{2m\omega ^{4}},
\end{equation}%
which is a time-dependent harmonic oscillator with a time-dependent free
term.

\section{Conclusions}

We have demonstrated that point transformations can be utilized to construct
non-Hermitian invariants for non-Hermitian Hamiltonians. In turn these
invariants may then be used to construct Dyson maps simply in form of
similarity transformations, which automatically satisfy the time-dependent
Dyson equation (\ref{TDDE}). Thus we have bypassed solving this more
complicated equation directly. When starting from an exactly solvable
reference Hamiltonian the scheme yields also the solution for the TDSE of
the target Hamiltonian. By construction the solutions only form an
orthonormal system when equipped with a metric operator that is obtained
trivially from the constructed Dyson map. We have shown that several
different reference Hamiltonians may lead to the same Dyson map.

It would be interesting to explore the scheme further by starting with more
complicated choices of the solvable reference Hamiltonian. However, the
scheme is of course not limited to exactly solvable models and we could also
start with a non-exactly solvable model as a reference system. In such a
setting the scheme would still yield an exact invariant and an exact metric
operator. Approximated wavefunctions could then be obtained by using the
procedure proposed in \cite{fring2020time}. Another interesting challenge is
to extend the scheme to higher dimensional systems.

\medskip

\noindent \textbf{Acknowledgments:} RT is supported by a City, University of
London Research Fellowship.

\newif\ifabfull\abfulltrue


\end{document}